\documentclass{article}
\usepackage{microtype}
\usepackage{graphicx}
\usepackage{subfigure}
\usepackage{booktabs} 
\usepackage{hyperref}

\usepackage[accepted]{icml2023}
\usepackage{amsmath}
\usepackage{amssymb}
\usepackage{mathtools}
\usepackage{amsthm}
\usepackage[capitalize,noabbrev]{cleveref}

\theoremstyle{plain}

\theoremstyle{definition}

\theoremstyle{remark}

\usepackage[textsize=tiny]{todonotes}

\icmltitlerunning{Interactively Optimizing Layout Transfer for Vector Graphics}

\begin{document}

\definecolor{darkgreen}{rgb}{0,0.5,0}
\definecolor{orange}{rgb}{1,0.5,0}
\definecolor{teal}{rgb}{0,0.5,0.5}
\definecolor{darkpurple}{rgb}{0.5, 0, 0.5}


\newcommand {\bjoern}[1]{{\color{blue}\bf{BH: #1}\normalfont}}
\newcommand {\jeremy}[1]{{\color{teal}\bf{JW: #1}\normalfont}}
\newcommand {\shuyao}[1]{{\color{darkpurple}\bf{SZ: #1}\normalfont}}


\newcommand {\add}[1]{{#1}}

\newcommand*{\quoted}[1]{{\small{\fontfamily{cmss}\selectfont{#1}}}}
\newcommand*{\participant}[1]{{\textbf{\small{\fontfamily{cmss}\selectfont{#1}}}}}
\newcommand{\etal}{et al. }

\newcommand {\bt}[1]{\textbf{#1} \normalfont}
\newcommand{\squishlist}{
 \begin{list}{$\bullet$}
  { \setlength{\itemsep}{0pt}
     \setlength{\parsep}{3pt}
     \setlength{\topsep}{3pt}
     \setlength{\partopsep}{0pt}
     \setlength{\leftmargin}{1.5em}
     \setlength{\labelwidth}{1em}
     \setlength{\labelsep}{0.5em} } }
\newcommand{\squishend}{
  \end{list}  }

\graphicspath{{./figures/}}

\twocolumn[
\icmltitle{Interactively Optimizing Layout Transfer for Vector Graphics}
\icmlsetsymbol{equal}{*}

\begin{icmlauthorlist}
\icmlauthor{Jeremy Warner}{eecs}
\icmlauthor{Shuyao Zhou}{eecs}
\icmlauthor{Björn Hartmann}{eecs}
\end{icmlauthorlist}

\icmlaffiliation{eecs}{EECS, UC Berkeley, Berkeley, CA, USA}

\icmlcorrespondingauthor{Jeremy Warner}{jeremy.warner@berkeley.edu}
\icmlkeywords{Vector Graphics, UI Design, Automatic Layout, Machine Learning, ICML}

\vskip 0.3in
]

\printAffiliationsAndNotice{}

\setlength{\footskip}{3.30003pt}

\begin{abstract}

Vector graphics are an industry-standard way to represent and share a broad range of visual designs.
Designers often explore layout alternatives and generate them by moving and resizing elements.
The motivation for this can range from establishing a different visual flow, adapting a design to a different aspect ratio, standardizing spacing, or redirecting the design's visual emphasis.
Existing designs can serve as a source of inspiration for layout modification across these goals.
However, generating these layout alternatives still requires significant manual effort in rearranging large groups of elements.
We present VLT, short for \emph{Vector Layout Transfer}, a novel tool that provides new techniques (Table~\ref{tab:techniques}) for transforming designs which enables the flexible transfer of layouts between designs. It provides designers with multiple levels of semantic layout editing controls, powered by automatic graphics correspondence and layout optimization algorithms.

\end{abstract}

\begin{figure}[t]
\vskip 0.2in
\begin{center}
\centerline{\includegraphics[width=\columnwidth]{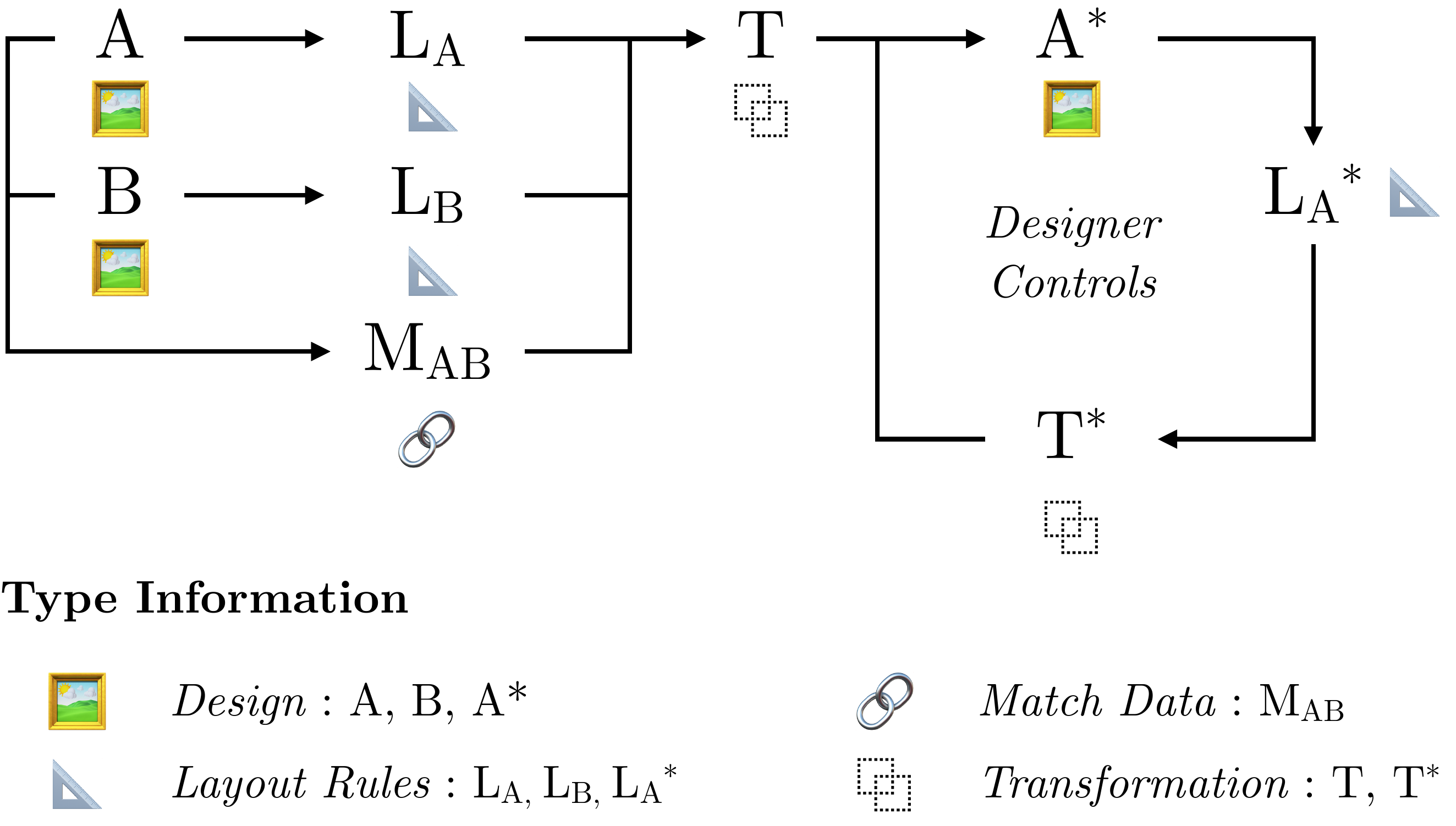}}
\caption{
Our layout transformation pipeline: given two vector graphics designs (A, B), we distill design layout data into grouped semantic layout rules for each design (L\textsubscript{A}, L\textsubscript{B}). 
We also compute a correspondence between the elements of the two designs (M\textsubscript{AB}).
Using L\textsubscript{A}, L\textsubscript{B}, and M\textsubscript{AB}, we generate T: a transformation of the graphic design elements of A.
Applying this transformation T yields design A*, which we then distill new layout rules from (L\textsubscript{A}*).
Designers can view the applied transformation and leverage control over which rules are prioritized, yielding new transformation T*, which in turn yields a new design.
This last component is an interactive, iterative process that aims to let designers retain full control of their design's layout while benefitting from automation.
}
\label{teaser}
\end{center}
\vskip -0.2in
\end{figure}

\section{Introduction}

Vector graphics are an industry-standard representation for many visual designs (e.g., logos, posters, interface mockups, and data visualizations).
Some artists choose to primarily work with vector graphics over other representations because vectors best suit curvilinear geometry and give `cleaner' aesthetics in their final result \cite{li2021artists}.
While this cleanliness and scalability are two reasons for vector graphics' success, another critical aspect is the flexibility of adapting layouts with discrete objects (vs. rasters).

Humans have both natural biological inclinations and learned heuristics for inferring information from a design element's scale, position, and ordering.
Perception of visual information is a well-established field, characterizing the different properties, aesthetics, and relations that objects can have to each other and what the effect is on the viewer \cite{bruno1988minimodularity, purchase1997aesthetic, card1999readings}.
Larger elements tend to capture more attention from viewers, and the relative arrangement and position of individual elements also influence the design's visual focus.
As a result, layouts are a core part of design in relation to attention and perception, ranging from map design \cite{wu2020survey}, data visualizations \cite{heer2010perception}, mobile user interfaces \cite{oyibo2020effect}, and more generally across graphic design \cite{zhao2018characterizes, bylinskii2017learning, fosco2020predicting}.
Skilled designers orchestrate these relational qualities, such as alignment, ordering, and sizing, to effectively allocate and streamline viewers' attention toward the key information they aim to convey.
This layout process is an iterative task involving resizing and moving many objects and possibly adding or removing content altogether.
Designers often explore the relational positions and layout of a vector graphics design to explore the effects of different variations \cite{samara2017making}.

Designers leverage many heuristics about what layout rules they should retain and which they should release to transform their designs.
Editing relational features like ordering, relative offsets, and alignment for different groups of objects is a bottleneck task in this design process that diminishes the designers ability to explore new designs.
While vector graphics are scalable, the relative dimensions (aspect ratio) and actual viewport size influence the preferred way to display information (e.g., mobile/desktop or poster/billboard), and reflowing an existing set of elements to a different size has been explored in related work \cite{hoffswell2020techniques}.

However, often the source of inspiration for wanting to change the layout of a design is not simply resizing but matching another design's layout; to \emph{transfer} the layout from a source or given example design.
Here, layouts are used to modify designs for greater purposes, including redirecting viewers' attention across the design and redistributing visual emphasis within the same design elements.
To facilitate this transfer of layouts across designs, we showcase a new tool (VLT) for vector graphic design layout transfer.
Our approach to this layout design transfer problem is to (a) infer and parameterize layout rules present in a given design and (b) facilitate the interactive transfer and iterative refinement of those rules via multiple levels of semantic editing.
We provide these varied levels of semantic editing and more powerful transformations with automatic graphics correspondence and layout optimization algorithms.

To enable \emph{layout transfer}, we extract relational rules from a given source design and apply those layout rules to a given target design.
This technique can reposition elements dynamically from a broad set of example designs.
Enabling transfer involves (a) inferring which relationships to retain vs. those which to break, (b) creating a correspondence between the two designs' elements to map adjustments across designs, and (c) computing and applying the minimal set of edits to integrate the source design's layout.

Our approach also involves iteratively refining and specifying how the layout is transferred with a range of techniques (Table\ref{tab:techniques}): (a) globally copying over layout rules for all elements, (b) copying all layout rules for a subset of elements, (c) specifying which rules design elements should adhere to, (d) specifying which properties to change per element, and finally (e) manually adjusting design elements with direct manipulation on the output canvas.
The set of rules (e.g., L\textsubscript{A}) for the output canvas updates in real time.

Our contributions include the following:
\textbf{(1)} a description of a pipeline for interactively optimizing layout transfer across designs;
\textbf{(2)} VLT, a novel tool that implements this pipeline;
\textbf{(3)} an gallery of example results generated with our tool.

\begin{figure*}[ht]
\vskip 0.2in
\begin{center}
\centerline{\includegraphics[width=\textwidth]{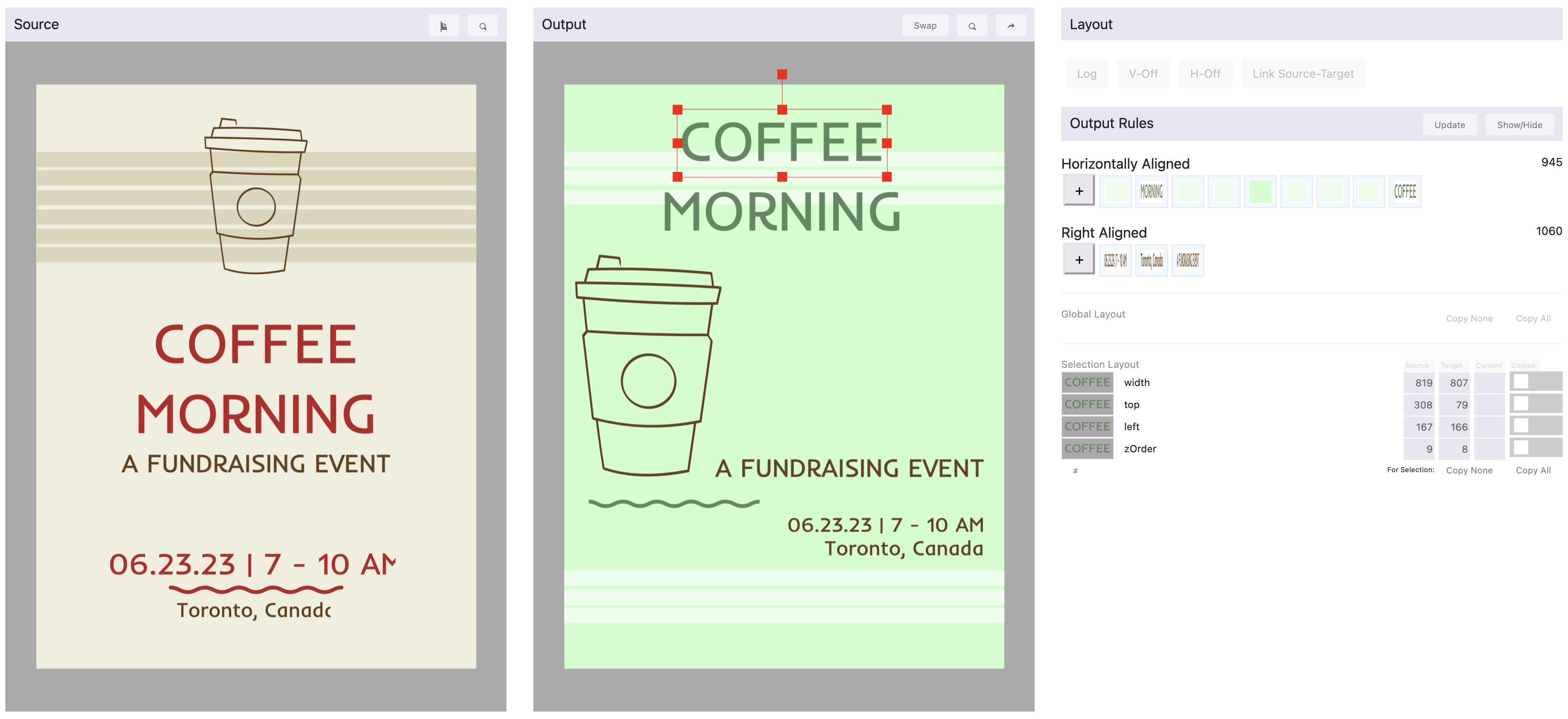}}
\caption{The VLT interface showing the source layout (e.g., B), the output layout (e.g., A*), and the layout rule customization panel. This output and the original target (A) can be toggled. The layout rules dynamically update as the output canvas is updated; here they show detected horizontal and right alignment rules. There are also global and element-specific layout transfer buttons, and a per-element property transfer based on that element's matched element.
This also works for multiple selected elements, grouping alike values.
}
\label{interface}
\end{center}
\vskip -0.2in
\end{figure*}

\begin{figure*}[ht]
\vskip 0.2in
\begin{center}
\centerline{\includegraphics[width=\textwidth]{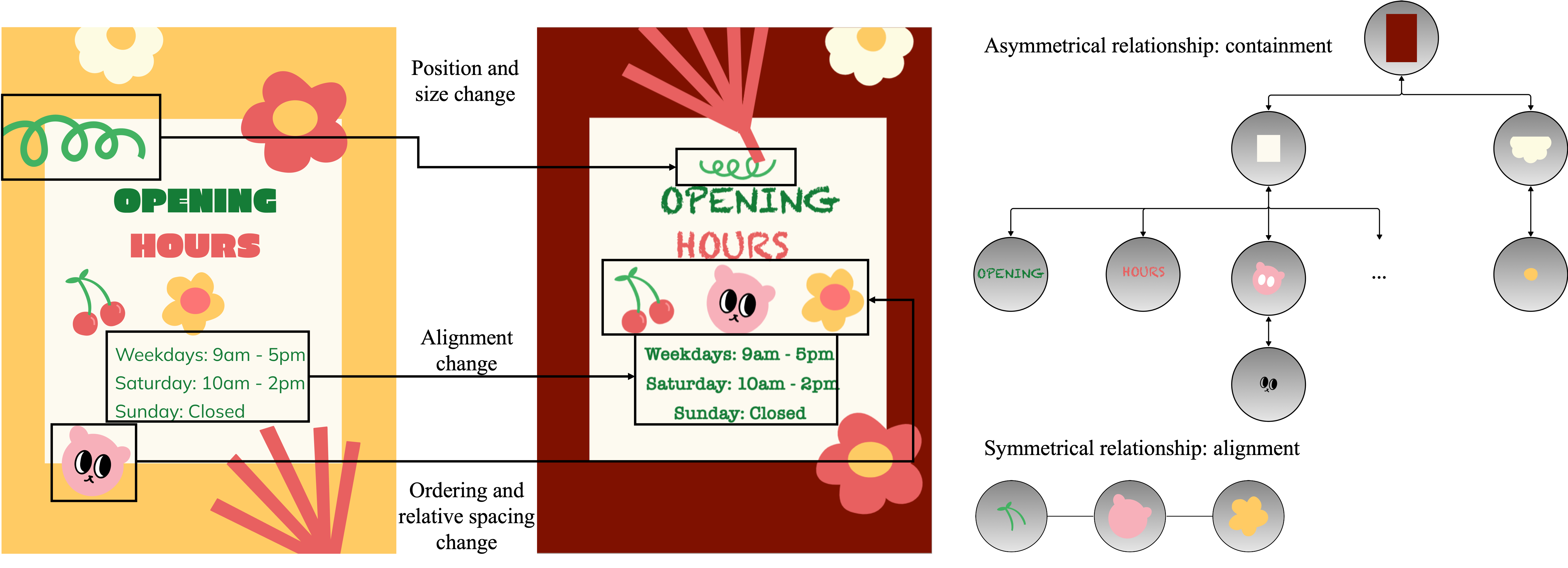}}
\caption{
The left side of this figure shows two designs with varying layouts, along with differing layout rules that were inferred for corresponding groups of elements.
The boxes and links in these designs represent different rule types that we recognize.
The right side shows a representation of the different types of layout relationships we can model between elements.
Asymmetric rules (e.g., containment) are represented internally as ordered trees while symmetric rules (e.g., alignment) are represented as simple sets (see also Table~\ref{tab:rules}).
}
\label{relations}
\end{center}
\vskip -0.2in
\end{figure*}

\section{Related Work}

We highlight two related areas: 
learning information about the patterns encoded in a given design and 
work that seeks to generate and manipulate the layouts of different designs.

\subsection{Design Patterns Recognition}
Recognizing design patterns plays a crucial role in a range of layout tasks.
In recent years, deep learning models have been proposed to address different aspects of vector graphics, including inference, generation, and editing \cite{ha2017neural, azadi2018multi, li2020differentiable, jain2022vectorfusion, ma2022towards, lawton2023drawing}. 
For UI design tasks specifically, previous research introduced a screen correspondence approach that focused on mapping interchangeable elements between user interfaces \cite{wu2023screen}.
This approach involves screen categorization and employs a UI Element Encoder to reason about the relative positions of UI elements. 
In the domain of UI layout understanding, the Spotlight model \cite{li2022spotlight} adopts a vision-language approach.
This model takes a combination of inputs, including a screenshot, a region of interest, and a text description of the task.
These inputs are processed using a multi-layer perceptron and a transformer model, resulting in a weighted attention output.
The output of the Spotlight model can be utilized for various UI design tasks (e.g., widget captioning, screen summarization).
Additionally, Shin et al. \cite{shin2021multi} proposed a multi-level correspondence method that leverages graph kernels to facilitate editing vector graphics designs.
This approach enables efficient editing of vector graphics by computing element-to-element correspondences at different levels of granularity.
Building upon these existing approaches, our work incorporates a graph-kernel based method for inferring objects and computing correspondences across canvases.
We can leverage the structural information of the designs to establish correspondences and perform efficient inference across multiple graphic designs.

\subsection{Layout Generation} 
Prior works have explored different approaches for layout generation and manipulation.
Datasets such as Rico \cite{deka2017rico} and WebUI \cite{wu2023webui} can be used for training probabilistic generative models of UI layouts.
Recent approaches explored transformer-based models in generating layouts \cite{lee2020neural,arroyo2021variational, kong2022blt}.
\add{With Im2Vec, researchers used differentiable rasterization to vectorize raster images and interpolate between them \cite{Reddy_2021_CVPR}.
Others learned implicit hierarchical representations for vector graphics to aid generation tasks, though they have focused on simpler designs (e.g., fonts) \cite{carlier2020deepsvg, lopes2019learned}.
}
\add{For layout transfer task, the Bricolage algorithm \cite{bricolage2011} employed a technique for generating mappings between Web pages by dividing them into significant regions and rearranging the elements to reflect parent-child relationships within a tree structure. However, it specifically focuses on HTML/CSS content and does not encompass visual layout transfer for vector graphics.}
\add{Also, the wealth of example website designs that Bricolage could leverage for training is comparatively scarce for vector graphics.} 

DesignScape provides users with layout suggestions, improving the efficiency of brainstorming designs \cite{ODonovan2015DesignScapeDW}.
Li et al. used the idea of Generative Adversarial Networks and proposed a differentiable wireframe rendering layer, specifically improving alignment sensitivity and better visibility of overlapping elements \cite{li2019layoutgan}. 
Ye et al. \cite{ye2020penrose} proposed Penrose that aimed to create mathematical diagrams using a layout engine that compiled code into layout configurations with the least overall energy while satisfying constraints and objectives.
Cheng et al. \cite{cheng2023play} presented a latent diffusion model PLay and conditioned on user input guidelines to generate UI layouts.
Chai et al. \cite{chai2023layoutdm} introduced the LayoutDM framework, which utilized a transformer-based diffusion model to generate layouts by representing each layout element using geometric and categorical attributes.
This model employed a conditional denoising diffusion probabilistic model to gradually map samples from a prior distribution to the actual distribution of the layout data.
Like Chai et al. \cite{chai2023layoutdm}, Naoto et al. \cite{inoue2023layoutdm} utilized diffusion models to generate layouts.
Additionally, Dayama et al. \cite{dayama2021interactive} proposed an interactive layout transfer system that allowed users to transfer draft designs to predefined layout templates.
Their approach relied on an interactive user interface and an existing library of layout templates.
However, the system required that the component types be predefined and rigidly categorized as either headings, containers, or content.
Our tool can transfer the user-input target layout onto the source design while retaining layout rules and consistency inferred from the designs, giving more flexibility for design tasks.

\section{VLT Walkthrough}
\label{walkthrough}

The broadest set of use cases for a tool like VLT is when designers would like to transform the layout of an existing design with a source reference design.
Figure~\ref{teaser} shows an overview of how designers can use VLT to transfer layouts across designs,
\add{and Table~\ref{tab:techniques} shows the core controls that VLT provides to designers for transforming the layout of their design using the source design as a source of inferred example rules.}
This walkthrough focuses on the iterative cycle designers can leverage to refine their output layout.

First, designers load two graphic designs A and B  into VLT (A = target = existing design to transform, and B = source = reference design).
Next, VLT will generate a correspondence matrix and match information (M\textsubscript{AB}) between the two sets of design elements \cite{shin2021multi}.
VLT also infers sets of semantic rules (listed in Table~\ref{tab:rules}) for each layout.

Designers can then copy the layout of the previous source design globally by inferring the position and size from the matched elements across designs.
The initial base transformation T uses the corresponding elements' base position and sizing, often giving subpar results (Figure~\ref{results}).
This naive case works on designs with a perfect one-to-one correspondence between design elements. 
However, many designs vary in the amount and type of elements they contain.
Designs may also change in their canvas size or aspect ratio, which copying position and size alone cannot address.

In these cases, VLT can be used to retain and adjust layout rules present in the original target design.
There is also an incremental rule-based optimization pipeline designers can leverage based on heuristic design rules (e.g., L\textsubscript{A}).
The dynamic set of layout rules that VLT infers can be viewed and modified in the right-most layout column of the interface (Figure~\ref{interface}), and a more detailed example with rule callouts is shown in Figure~\ref{relations}.
The rule list updates according to the selected canvas elements.
This brings the designers' attention to controls for leveraging these rules to modify their designs' layouts.
Elements may be manually adjusted (i.e., direct manipulation) on the output canvas, and the set of detected layout rules updates in real time.
\begin{table}[h]
    \centering
    \caption{Designer Controls for Layout Editing}
    \begin{tabular}{ll}
        & \\
         \textbf{Granularity} & \textbf{Technique} \\
         \hline
        Highest & Global Layout Copy \\
        & Element Layout Copy \\
        & Individual Rule Adherence \\
        & Correspondence Update \\
        & Element Property Copy \\
        Lowest & Direct Manipulation \\
    \end{tabular}
    \label{tab:techniques}
\end{table}
\begin{table}[h]
    \centering
    \caption{Supported Layout Heuristic Rules (e.g., L\textsubscript{A})}
    \begin{tabular}{ll}
            & \\
         \textbf{Type} & \textbf{Name} \\
         \hline
        Asymmetric & Containment \\
        & Relative Ordering \\
        Symmetric & V/H Alignment \\
        & Bounds Overlap \\
        & Marginal Offset \\
        & Same W/H \\
    \end{tabular}
    \label{tab:rules}
\end{table}
In addition to copying the layout of an entire design, designers may opt only to transfer (or reset) layout properties for specifically selected elements.  
Other elements can be added from layout rules here (clicking the \texttt{+} next to the rule member list) and conforming the marginal spacing across design versions.
For example, selecting the H-Off or V-Off buttons will adjust the marginal spacing and offset for the currently selected elements to an inferred value based on their match.
Designers may select elements from the source design (B), observe the rules they adhere to, and apply them (or a rule subset) to elements on the output canvas.
Once satisfied, they can export the transformed design as an SVG.

\section{Optimizing Layouts}
\label{optimize}

To optimize the transferral of a layout across designs, we must first create a representation of that layout.
We construct a transformation T that includes scale and translation amounts per graphic element to do this.
Similarly, we first represent the layout of a specific visual design A as the \emph{position} and \emph{size} of each graphical element ($e$).

\begin{equation}
    e \to [x, y, z, w, h]
\end{equation}

Note that $z$ here refers to the z-index or relative layering, while $x$ and $y$ refer to the uppermost, leftmost element canvas point for that element.
Also, $w$ and $h$ refer to the element's canvas width and height, respectively.
So, a given transformation T to transform a graphic design A would consist of a set of changes to these element properties:

\begin{equation}
    T \to \forall e \in A : [\delta x, \delta y, \delta z, \delta w, \delta h]  
\end{equation}

On top of this broad representation, we also build up sets of heuristic-based rules (e.g., L\textsubscript{A}, L\textsubscript{B}) that we can relate across multiple designs.
These rules include containment, ordering, alignment, overlapping elements, relative margins, and size (Table~\ref{tab:rules}), which may have either symmetric or asymmetric relations between elements.
For example, alignment is symmetric in that all elements have the same relationship with each other (internally represented in VLT as a set), while containment has a structured ordering between related elements (internally represented as an ordered tree).
Visual examples of the distinction between symmetric and asymmetric rules are shown in Figure~\ref{relations}.

The optimal T choice for an exact one-to-one pairing of design elements is obvious -- rescale and reposition the elements precisely to where they were in the corresponding design. 
However, there clearly are better ways to edit graphics than manually adjusting x and y coordinates.
Recognizing and leveraging inferred design rules is a promising direction toward using automation while retaining designer control.
We also want to handle complex one-to-many mappings between the sets of design elements.

First, layout rules from the source for corresponding elements are applied to the output graphics.
This is initially done using the matched element's position and size, which may cause multiple elements to overlap (Figure~\ref{results}).
To alleviate this, we also provide buttons to extend the marginal offset (Vertical-Offset/Horizontal-Offset) between matched elements onto the linked target elements.
Individual rules can be specified to recompute a transformation that complies with the specified rule.
This iterative optimization is an active project development area, and we detail ongoing work in our layout optimization in \ref{future}.

\begin{figure}[t]
    \vskip 0.2in
    \begin{center}
    \centerline{\includegraphics[width=0.5\textwidth]{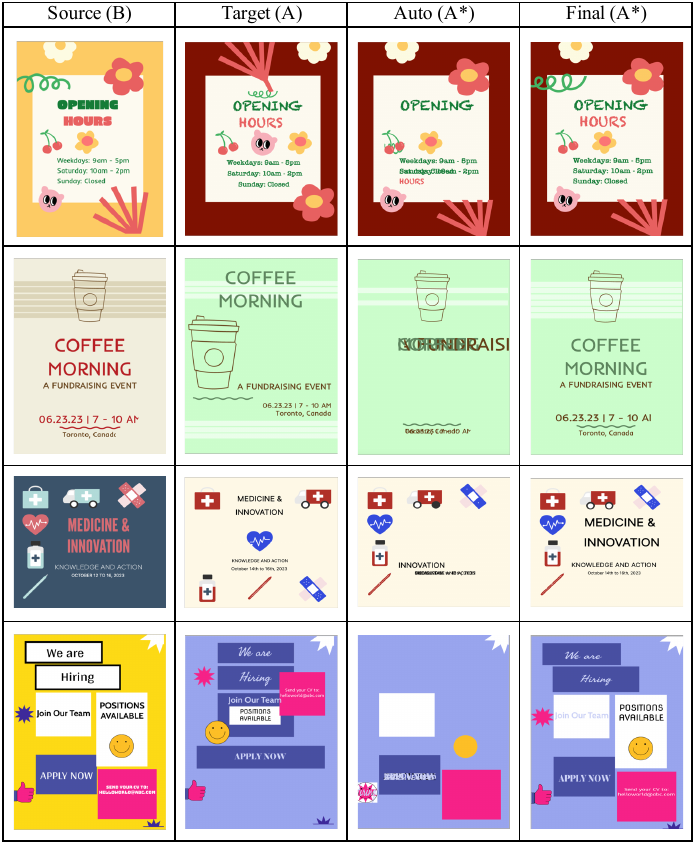}}
    \caption{
    Example output gallery of layouts made with VLT.
    Each column shows (in order from columns 1-4): the source or inspiring layout (B), the target input design (A), the fully automatic result of globally applying layout transformation rules to the entire design with no iterative designer control, and the final output design iteratively made with VLT's range of semantic editing controls.
    }
    \label{gallery}
    \end{center}
    \vskip -0.2in
\end{figure}

\section{Design Results}
\label{results}
To showcase the effectiveness of our method, we provide several example graphics that were transformed using the pipeline and tool detailed in this paper in Figure~\ref{gallery}.
The generation of these graphics was done by the authors using VLT. 
We aim to include more complex and varied examples, and have actual designers use VLT to transfer layouts across existing designs.
For the graphics we generated, the amount of UI interactions to transform each design from Target to Final (per row) is 7 / 8 / 12 / 15, and the total number of transformed element properties is 111 / 76 / 291 / 128.  
The higher numbers for the property changes reflect that many properties can be changed with a single UI interaction in VLT.
The procedure we followed to transfer layouts was to first match designs, transfer the global layout using the correspondence, leverage layout rules as needed, and finally tweak elements directly on the canvas.
This follows granularity shown in Table~\ref{tab:techniques}; paint with the broadest strokes initially and iteratively handle smaller outlier classes.

\section{Discussion}
\label{discussion}

We discuss two main topics:
(1) reflections from balancing designer control with boosting editing workflows with automation, and
(2) limitations of working with layouts in this way and future steps we envision taking to address this.

\subsection{Balancing Control \& Automation}
As automation-driven media creation and manipulation tools proliferate, there is a valid concern about displacing the designer from their current creative control.
Our goal in this project is to retain the final control that designers have over their designs while reducing some of the tedium and manual labor that goes towards manifesting a specific vision for that given design.
Our high-level approach towards this goal involves sharing a range of dynamic controls that the designer can adapt to the level of detail they wish to edit at, a sort of semantic range of design detail to operate over.

One of the ways we aim to provide this balance of control and automation includes providing several levels of detail and forms of editing and specifying transformation rules with VLT.
This approach includes displaying inferred layout rules that can also modify existing designs,
displaying editable global and element-specific layout data, 
and enabling live updates as the designer modifies their output (including via direct manipulation).
Generally, the more deeply intertwined any automation becomes into existing creative practices necessitates deeper robustness and reliability to successfully operate `as expected', which for many domains (image style transfer, text-to-image creation, vector layout transfer) remains a challenging and subjective task.

\subsection{Limitations \& Future Work}
\label{future}

\emph{Layout Optimization.}
The current process for initially learning a layout transformation T is driven by correspondences, then refined by leveraging manually-crafted design heuristics.
We want to leverage a more flexible approach to both initially craft and incorporate designer demonstrations and updates into design layout transformations.
We envision using a combination of heuristic layout information currently gleaned from the SVG canvas and other vision-based UI understanding features to bolster the layout transformation and optimization process.
Additionally, our current design transformation only consists of rescaling (height, width) and repositioning (x, y, z/layer) design elements.
Other valid transformations exist, such as rotation and skew, but we have yet to implement them as we have found them less common.
Enabling these transformations may yield additional desired variations that VLT cannot currently produce.
\add{
We also take inspiration from \cite{bricolage2011}, which details a technique for learning the cost of connecting edges across sets of web design elements.
They infer a new semantic hierarchy tree for both designs and compute a minimal cost mapping across the nodes of the trees.
To do this, they train a standard multi-layer perceptron for optimizing weights related to retaining tree ancestry, node siblings, and unmatched elements.
This optimization also considers the visual and semantic properties of each node that they match.
They base their training on a set of human-provided mappings across visual design examples. 
Also, the optimization in their work focuses on producing a mapping between design elements, while we seek to optimize a transformation of one design's layout based on that mapping, compared to the mapping itself.
}

\emph{Differentiable Layouts.}
Adherence to a discrete set of recognized layout rules is difficult to optimize because of the binary nature of rule groups -- elements either adhere or not.
To enable optimization of this discrete model, we are working to build a reward function $R_T$ for transformation T based on the \emph{relative} adherence and weight of inferred design heuristics and rules.
We will apply Gaussian smoothing to the position and width/height constraints for symmetric relations like alignment, element overlap, offset, and sizing (Table~\ref{tab:rules}). 
Here, $r$ represents the layout rules that applying T yields, $\omega _i$ is the rule weight (which designers may adjust in a range of ways), and $e _r$ measures how many elements correspond to that rule.

\begin{equation}
\begin{split}
    R_T &= R_{\text{rule}} + R_{\text{off}} + R_{\text{con}} \\
    R_{\text{rule}} &= \sum_{r} \omega _r * log(e _r + 1) \\
    R_{\text{off}} &= \omega _{\text{off}} * t_{\text{non-overlap}} \\
    R_{\text{con}} &= \frac{\omega _{\text{con}}}{ e_{\text{unique-prop}}} 
\end{split}
\end{equation}

In addition to general rule adherence, we propose metrics $R_{\text{off}}$ for balancing the relative offset of objects (e.g., favor non-occlusion of text) and $R_{\text{con}}$ for increasing the numeric consistency of almost-alike element properties, a sort of snap-to-fit implementation (e.g., favor sizing/spacing).
Also, $t_{\text{non-overlap}}$ refers to the non-overlapping text elements, and $e_{\text{unique-prop}}$ refers to the number of unique properties that exist in a design (less is better).
These rewards also will global adjustable weights ($\omega _{\text{off}}$, $\omega _{\text{con}}$), respectively.
 
Designers will be able to selectively apply this optimization to part of the design or simply run it over the entire output design.
In addition, we can optimize specific inferred rules from the source or target while retaining as much structure from the alternative goals as possible by explicitly increasing the weight of those sections.
Designers could opt to lock constrained element properties in their design (e.g., size) to ensure those properties are not modified, or extend a manually demonstrated layout change to similar elements.

\emph{Element Correspondence.}
When designs have elements that are alike, finding a correspondence between the two element sets (M\textsubscript{AB}) is natural.
However, this element correspondence between designs will often be noisier or less accurate for very unrelated or larger designs.
One direction for future work we envision is being able to dynamically infer a set of joint classes across elements, of which design elements might belong to many, as opposed to a cross-design element map.
VLT shows grouped layout rules and property changes, but the level of inference could be smoother and capture a broader set of similarities to enhance designer control.

\emph{User Evaluation.}
To measure the effectiveness of our approach to modifying vector graphics designs, we would like to work with actual designers to see how they might leverage this tool, whether they could incorporate it into their existing workflows, and what changes would make it truly useful to them.
\add{The main evaluation metrics would be
\textbf{(1)} what the comparative difficulty and timing would be for producing designs from the same prompt, and
\textbf{(2)} subjective quality ratings to see if someone using VLT can make similar quality layout transformations compared to expert designers.}
We plan on running a user study with experienced designers where they will use VLT to transform several graphics design layouts to enhance the layout transfer process and enhance consistency among designs.
\add{We envision that designers using VLT could create high-quality layout designs in less time than when creating designs with traditional manual vector graphics editing software.}

\emph{Technical Evaluation.}
\add{Another way of measuring the effectiveness of VLT will be to evaluate the quality of a fully automatic approach to transferring design rules.
Like Bricolage ~\cite{bricolage2011}, we could leverage human examples to work towards this automation. 
Instead of collecting human-constructed element mappings, we could recruit designers to transfer layouts across designs as training examples.
We envision a technical evaluation to characterize our approach that would leverage the number of VLT UI interactions and individual property edits to get from an initial automatic set of transfer results to our human-provided goal layout.
We could also share this example set as a benchmark for progressing on this challenging vector layout transfer task.
}

\section{Conclusion}
\label{conclusion}

Our paper presents a novel design tool, VLT, that can enable interactive layout transfer optimization.
VLT's process for inferring and transferring layouts (Figure~\ref{teaser}) integrates automation into the design process while providing several levels of automation-driven semantic control and editing techniques (Table~\ref{tab:techniques}) for designers to steer and adjust the resulting final layout.
We showcase some preliminary results (Figure~\ref{gallery}) and highlight several important next steps for addressing the broader challenge of layout transfer.

\bibliography{paper}
\bibliographystyle{icml2023}

\end{document}